\documentclass{aa}  

\usepackage{graphicx}
\usepackage{txfonts}
\usepackage[]{hyperref} % To add links in your PDF file
\usepackage{booktabs} % for much better looking tables
\usepackage{array} % for better arrays (eg matrices) in maths
\usepackage{multirow} % For multirow cells
\usepackage{paralist} % very flexible & customisable lists (eg. enumerate/itemize, etc.)
\usepackage{verbatim} % adds environment for commenting out blocks of text & for better verbatim
\usepackage{ulem}
\usepackage{xspace}
\usepackage{siunitx}
\usepackage{amssymb}
\usepackage{capt-of}
\usepackage{natbib}
\bibpunct{(}{)}{;}{a}{}{,} % to follow the A&A style
\usepackage{float}
\usepackage{placeins}
\graphicspath{{./img/}}

\makeatletter
\renewcommand*\aa@pageof{, page \thepage{} of \pageref*{LastPage}}
\makeatother

\addto\extrasenglish{%
}

\newcommand\freefootnote[1]{%
  \let\thefootnote\relax%
  \footnotetext{#1}%
  \let\thefootnote\svthefootnote%
}

\begin{document}

   \title{Retrieving wind properties from the ultra-hot dayside of WASP-189\,b with CRIRES$^+$}

   \subtitle{}

   \author{
            F. Lesjak \inst{1}\thanks{fabio.lesjak@uni-goettingen.de}\and 
            L. Nortmann \inst{1} \and
            D. Cont \inst{2,3}  \and
            F. Yan \inst{4} \and
            A. Reiners\inst{1} \and
            N. Piskunov\inst{5} \and
            A. Hatzes\inst{6} \and
            L. Boldt-Christmas\inst{5}\and
            S. Czesla\inst{6} \and
            % U. Heiter\inst{5} \and
            % O. Kochukhov\inst{5} \and
            A. Lavail\inst{7} \and
            E. Nagel\inst{1} \and
            A. D. Rains\inst{5} \and
            M. Rengel\inst{8} \and
            % F. Rodler\inst{10} \and
            U. Seemann\inst{9, 1}\and
            D. Shulyak \inst{10}
          }

   \institute{
            Institut f\"ur Astrophysik und Geophysik, Georg-August-Universit\"at, Friedrich-Hund-Platz 1, 37077 G\"ottingen, Germany \and
            Universitäts-Sternwarte, Fakultät für Physik, Ludwig-Maximilians-Universität München, Scheinerstr. 1, 81679 München, Germany \and
            Exzellenzcluster Origins, Boltzmannstraße 2, 85748 Garching, Germany \and
            Department of Astronomy, University of Science and Technology of China, Hefei 230026, China \and
            Department of Physics and Astronomy, Uppsala University, Box 516, 75120 Uppsala, Sweden \and
            Thüringer Landessternwarte Tautenburg, Sternwarte 5, 07778 Tautenburg, Germany \and
            Institut de Recherche en Astrophysique et Planétologie, Université de Toulouse, CNRS, IRAP/UMR 5277, 14 avenue Edouard Belin, F-31400, Toulouse, France \and
%             Hamburger Sternwarte, Universität Hamburg, Gojenbergsweg 112, 21029 Hamburg, Germany \and
            Max-Planck-Institut für Sonnensystemforschung, Justus-von-Liebig-Weg 3, 37077 Göttingen, Germany \and
%             European Southern Observatory, Alonso de Cordova 3107, Vitacura, Santiago, Chile \and
            European Southern Observatory, Karl-Schwarzschild-Str. 2, 85748 Garching bei München, Germany \and
            Instituto de Astrofísica de Andalucía - CSIC, Glorieta de la Astronomía s/n, 18008 Granada, Spain
             }

   \date{Received 5 July 2024; accepted 21 November 2024}

% \abstract{}{}{}{}{} 
% 5 {} token are mandatory
    % \abstract{Text here}
  \abstract
  % context heading (optional)
  % {} leave it empty if necessary  
   {The extreme temperature gradients from day- to nightside in the atmospheres of hot Jupiters generate fast winds in the form of equatorial jets or day-to-night flows. Observations of blue-shifted and red-shifted signals in the transmission and dayside spectra of WASP-189\,b have sparked discussions about the nature of winds on this planet.}
  % aims heading (mandatory)
   {To investigate the structure of winds in the atmosphere of the ultra-hot Jupiter WASP-189\,b, we studied its dayside emission spectrum with CRIRES$^+$ in the spectral K band. }
  % methods heading (mandatory)
   {After removing stellar and telluric lines, we used the cross-correlation method to search for a range of molecules and detected emission signals of CO and Fe. Subsequently, we employed a Bayesian framework to retrieve the atmospheric parameters relating to the temperature-pressure structure and chemistry, and incorporated a numerical model of the line profile influenced by various dynamic effects to determine the wind structure.}
  % results heading (mandatory)
   {The cross-correlation signals of CO and Fe showed a velocity offset of $\sim$6\,km\,s$^{-1}$, which could be caused by a fast day-to-night wind in the atmosphere of WASP-189\,b. The atmospheric retrieval showed that the line profile of the observed spectra is best fitted by the presence of a day-to-night wind of $4.4^{+1.8}_{-2.2}$\,km\,s$^{-1}$, while the retrieved equatorial jet velocity of $1.0^{+0.9}_{-1.8}$\,km\,s$^{-1}$ is consistent with the absence of such a jet. Such a wind pattern is consistent with the observed line broadening and can explain the majority of the velocity offset, while uncertainties in the ephemerides and the effects of a hot spot could also contribute to this offset. We further retrieved an inverted temperature-pressure profile, and under the assumption of equilibrium chemistry we retrieved a C/O ratio of $0.32^{+0.41}_{-0.14}$ and a metallicity of [M/H] = $1.40^{+1.39}_{-0.60}$.}
  % conclusions heading (optional), leave it empty if necessary 
   {We showed that red-shifts of a few km\,s$^{-1}$ in the dayside spectra could be explained by day-to-night winds. Further studies combining transmission and dayside observations could advance our understanding of WASP-189\,b's atmospheric circulation by improving the uncertainties in the velocity offset and wind parameters.}


   \maketitle

\section{Introduction}

Ultra-hot Jupiters (UHJs) are a class of highly irradiated gas giant planets with equilibrium temperatures exceeding 2000\,K. The daysides of these planets are expected to be free of clouds \citep[e.g.][]{helling2021CloudPropertyTrends}, which makes them favourable targets for emission spectroscopy. The extreme temperatures lead to molecular dissociation and ionisation of atoms \citep{parmentier2018A&A...617A.110P, helling2019A&A...626A.133H}, and the stark day-night temperature gradient creates fast winds \citep[e.g.][]{Showman2002A&A...385..166S, Rauscher2010ApJ...714.1334R}.

Simulations of UHJ atmospheres predict the formation of zonal equatorial jets in the direction of the planet's rotation, along with day-to-night winds transporting material across the terminator \citep[e.g.][]{zhang2017ConstrainingHotJupiter}. The intensity of these winds depends on a variety of factors, such as the equilibrium temperature, rotation period, and frictional drag \citep{tan2019ApJ...886...26T}. High-resolution spectroscopy can be used to study the radial velocities produced by these effects, such as velocity shifts due to day-to-night winds \citep[e.g.][]{Flowers2019AJ....157..209F, Seidel2023A&A...673A.125S} and line broadening due to jets \citep[e.g.][]{cont2022AtmosphericCharacterizationUltrahot, lesjak2023A&A...678A..23L}.

The ultra-hot Jupiter WASP-189\,b \citep{anderson2018WASP189bUltrahotJupiter} is thus far one of the hottest planets with a characterised atmosphere ($T_\mathrm{eq} = 3353$\,K, \citealt{lendl2020HotDaysideAsymmetric}). It orbits a bright A-type star with a period of 2.7\, days, and has been previously observed both from space with the CHaracterising ExOPlanets Satellite \citep[CHEOPS, ][]{lendl2020HotDaysideAsymmetric} and the Transiting Exoplanet Survey Satellite \citep[TESS, ][]{saha2023PreciseTransitPhotometry}, as well as from the ground using the High Accuracy Radial velocity Planet Searcher (HARPS) and HARPS-N \citep{yan2020TemperatureInversionAtomic, prinoth2021TitaniumOxideChemical, stangret2022HighresolutionTransmissionSpectroscopy} and GIANO \citep{yan2022DetectionCOEmission}. Near-ultraviolet observations suggest that the upper atmosphere can reach temperatures of 15\,000\,K \citep{Sreejith2023ApJ...954L..23S}. The stellar and planetary parameters can be found in Table \ref{Table_Parameters}. \citet{prinoth2021TitaniumOxideChemical} investigated the transmission spectrum in the visible wavelength region, detecting the presence of TiO and various metals like chromium, iron, and magnesium. They reported varying velocity offsets from the expected planetary rest frame for different species, suggesting different three-dimensional distributions caused by thermo-chemical stratification. A follow-up study \citep{prinoth2023TimeresolvedTransmissionSpectroscopy} revealed that the signal strength of individual species varies over the duration of the transit, with most species showing increased absorption closer to egress. However, some species (e.g. Ni and TiO) were instead found to decrease in strength over time, which could be attributed to ionisation and dissociation on the planet's dayside.

The emission spectrum of WASP-189\,b was studied in detail by \citet{yan2020TemperatureInversionAtomic}, detecting Fe and constraining the temperature-pressure ($T$-$p$) profile, and \citet{yan2022DetectionCOEmission} found emission of CO in the near-infrared (NIR) observations with a significant red-shift of $4.5$\,km\,s$^{-1}$. This radial velocity offset could be induced by winds from the dayside to the nightside. However, a general circulation model of WASP-189\,b simulated by \citet{lee2022MantisNetworkII} resulted in fast zonal equatorial winds with speeds of 5-6\,km\,s$^{-1}$ in pressure regions from 10$^{-2}$ to 10$^{-4}$\,bar, and only a weak day-to-night wind. The $T$-$p$ profile showed a temperature inversion at longitudinally dependent heights ranging from 10$^{-5}$\,bar on the nightside to 10$^{-1}$ close to the sub-stellar point.

In this work, we present high-resolution dayside observations of WASP-189\,b with the CRyogenic InfraRed Echelle Spectrograph$^+$ (CRIRES$^+$, \citealt{dorn2014CRIRESExploringCold, dorn2023CRIRESSkyESO}). This instrument is the upgraded version of the original CRIRES \citep{kaeufl2004CRIRESHighresolutionInfrared} and enables high signal-to-noise (S/N) observations with sufficient time resolution and a spectral resolution of $R{\sim}100\,000$.
We retrieve the atmospheric properties of WASP-189\,b using a Bayesian retrieval framework, and analyse the spectral line shape to determine whether the atmospheric dynamics are predominantly shaped by day-to-night winds or an equatorial jet.

In Sect. \ref{Section_Data} we describe the observations and our data reduction procedure, and in Sect. \ref{Sect_CrossCorrelation} we detail the model generation and cross-correlation analysis. We present our retrieval analysis in Sect. \ref{Sect_Retrieval}, determine the jet and day-to-night wind speeds by applying a more sophisticated line profile model in Sect. \ref{Sect_WindRetrieval}, and conclude in Sect. \ref{Section_Conclusion}.

\begin{table}[ht]\renewcommand{\arraystretch}{1.5}
 \caption[]{Stellar and planetary parameters of the WASP-189 system.}\label{Table_Parameters}
\begin{tabular}{lll}
 \hline \hline
  Parameter &
  Symbol &
  Value

 \\ \hline
\textit{Planet}   &   & \\
Radius$^a$   & $R_\mathrm{p}$ ($R_\mathrm{Jup}$) &  1.619 $\pm$ 0.021  \\
Mass$^a$   & $M_\mathrm{p}$ ($M_\mathrm{Jup}$) &  1.99 $^{+0.16}_{-0.14}$  \\
Orbital period$^b$   & $P_\mathrm{orb}$ (days) & 2.7240338 \\%$\pm 0.0000042$\\
Orbital inclination$^a$    & $i$ ($^\circ$) &  84.03 $\pm$ 0.14 \\
Orbital eccentricity$^a$   & $e$ & 0 \\
Semi-major axis$^a$   & $a$ (AU)& 0.05053 $\pm$ 0.00098 \\
Time of mid-transit$^a$   & $T_0$ (BJD) & 2458926.541696 \\%$\pm ?$\\
RV semi-amplitude$^c$   & $K_p$ (km\,s$^{-1}$) & 201 $\pm$ 4 \\
Surface gravity$^c$   & $\log{g}$ (cgs)& 3.29 \\
Equil. temperature$^a$   & $T_\mathrm{eq}$ (K)& 3353 $^{+27}_{-34}$  \\
 \\ \hline
\textit{Star}   &   & \\
Radius$^a$   & $R_\star$ ($R_\odot$) & 2.36 $\pm$ 0.03  \\
Effective temperature$^a$   & $T_\mathrm{eff}$ (K)& 8000 $\pm$ 80\\
Systemic velocity$^b$   & $v_\mathrm{sys}$ (km\,s$^{-1}$) & $-24.452$ 
 \\ \hline
\end{tabular}
\tablebib{
$^{(a)}$\citet{lendl2020HotDaysideAsymmetric},
$^{(b)}$\citet{anderson2018WASP189bUltrahotJupiter},
$^{(c)}$ Calculated from orbital parameters: $K_\mathrm{p} = 2\,\pi\,a\,\sin(i)\,P^{-1}\,(1-e^2)$

}
\end{table}

\section{Observations and data reduction} \label{Section_Data}
\subsection{Observations}
We used the CRIRES$^+$ instrument at the 8m-class Unit Telescope 3 of the Very Large Telescope to observe WASP-189\,b on 01 July 2022 as part of the guaranteed time observations of the CRIRES$^+$ consortium. In order to detect the planetary dayside emission, the observation covered a part of the orbit shortly after the secondary eclipse. 
We used the K2148 wavelength setting, covering the spectral range from 1972\,nm to 2452\,nm split over six spectral orders. This wavelength range includes spectral lines of the major species expected in UHJ atmospheres (e.g. CO, H$_2$O, Fe). We chose an exposure time of 300\,s to ensure sufficient S/N per exposure while not smearing the planetary spectral lines over many detector pixels, as this effect has been proven to be impactful on high-resolution observations \citep{BoldtChristmas2024A&A...683A.244B}. CRIRES$^+$ is an echelle spectrograph with a science detector array comprised of 3 chips, so that each of the six spectral orders is split into three parts, resulting in 18 segments.

Nodding was employed during the observation, wherein the target position alternates between two different locations along the slit (A and B) with four consecutive exposures per nodding position. Information of both positions are combined during the extraction process to remove the background sky signal and detector artefacts. We treated the reduced spectra of A and B as two distinct sets of data, and applied the data processing and analysis on each time-series individually before combining the results during the cross-correlation.

To facilitate the wavelength calibration, metrology was employed prior to the science observations to fine-tune the positions of specified emission lines from Kr- and Ne-lamps. We chose a 0.2$\arcsec$ slit width to ensure a high nominal spectral resolution of $R{\sim}100\,000$. However, in our observation the adaptive optics system performed very well and the point-spread function (PSF) did not illuminate the whole slit, leading to an even higher resolution \citep{dorn2023CRIRESSkyESO}. We measured the width of the PSF to estimate the true resolution of $R{\sim}140\,000$, following the methodology of \citet{lesjak2023A&A...678A..23L}.

We reduced the raw spectra with the European Southern Observatory (ESO) CRIRES$^+$ data reduction pipeline \texttt{cr2res} (version 1.3.0), which includes dark and flat field corrections, bad pixel removal, and a derivation of the wavelength solution\footnote{The reduced data can be found at \url {https://doi.org/10.5281/zenodo.12663686} \citep{lavail_2024_12663686}}.
We further applied \texttt{molecfit} \citep{smette2015MolecfitGeneralTool} to the time-averaged spectrum of A and B individually to refine the wavelength solutions for both nodding positions. A further correction of the wavelength solution for each individual exposure was not necessary, as the deviations determined via cross-correlating the telluric lines (see \citealt{Cont2024arXiv240608166C}) were sufficiently small.

We discarded 40 pixels at the beginning and end of each segment to avoid detector edge effects, and masked regions that were heavily contaminated by telluric absorption (1997\,nm -- 2015\,nm, 2162\,nm -- 2170\,nm, 2314\,nm -- 2318\,nm, and 2450\,nm -- 2453\,nm).

\subsection{Normalisation}
We removed outliers in the time series of spectral bins with a 5$\sigma$-clipping, and corrected for the blaze function by following the three-step process of \citet{gibson2022RelativeAbundanceConstraints} to minimise the distortion of the planetary signal. First, the spectra of each segment were divided by the segment-wide median spectrum. Next, the resulting residual spectra were smoothed by applying a median filter of 501 pixels and then a Gaussian filter with a standard deviation of 100 pixels. Finally, each original spectrum was divided by the smoothed residual spectrum of the corresponding segment. This process removed time-dependent variations and brought all spectra on the same blaze function, which was then removed during the following filtering step. We further excluded deep telluric lines by masking the wavelength regions where the flux falls below 40\,\% of the continuum level.

\subsection{\texttt{SYSREM}}
\begin{figure}
\centering
\includegraphics[width=\hsize]{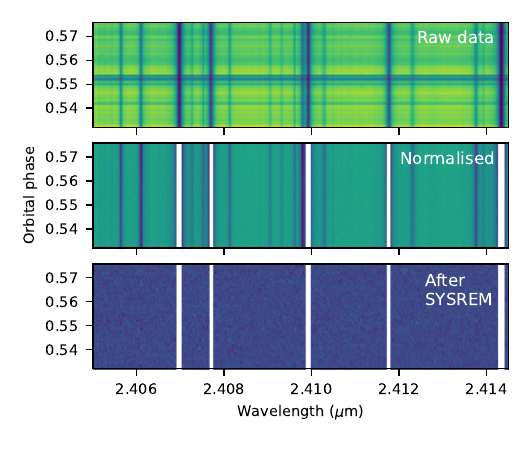}
  \caption{Data reduction steps of a representative wavelength range. Top panel: Unprocessed spectra as they are produced by the CRIRES$^+$ pipeline. Middle panel: After normalisation to bring all spectra onto a common blaze function, and masking of deep telluric lines. Bottom panel: After the removal of stellar and telluric lines with \texttt{SYSREM}.
          }
     \label{Fig_DataReduction}
\end{figure}

We employed \texttt{SYSREM} \citep{tamuz2005CorrectingSystematicEffects, birkby2013DetectionWaterAbsorption} to remove the blaze function and the telluric and stellar lines in the spectra, as detailed for example in \citet{gibson2020DetectionFeAtmosphere} and \citet{lesjak2023A&A...678A..23L}. \texttt{SYSREM} utilises a principal component analysis to construct a model of the linear components in wavelength and time, while incorporating the uncertainty of each data point. The observed spectra contain stellar, telluric, and planetary components with distinct Doppler shifts. While stellar and telluric lines remain nearly static in velocity space throughout the observation, the planetary radial velocity changes significantly. A comprehensive description of \texttt{SYSREM} in the context of high-resolution exoplanet observations can be found in \citet{Czesla2024A&A...683A..67C}.
We followed the method of \citet{gibson2022RelativeAbundanceConstraints} and first divided both the spectra and uncertainties by the time-averaged spectrum of each segment. Then we used \texttt{SYSREM} to iteratively subtract linear trends in time from each spectral pixel, effectively eliminating the (quasi-)static components of the spectra. This process resulted in residuals mainly comprised of noise and the planetary signal, which shifts in wavelength over time and is not modelled by \texttt{SYSREM}. The linear model of each iteration was refined until the average relative change fell below 0.01. The model was then subtracted from the data before beginning with the subsequent iteration. An example of the resulting residual spectra is shown in Fig. \ref{Fig_DataReduction}.
The number of \texttt{SYSREM} iterations impacts the recovered signal strength \citep{BoldtChristmas2024A&A...683A.244B}, and our methodology for determining this number is outlined in Sect. \ref{Sysrem iterations}. 

\section{Cross-correlation} \label{Sect_CrossCorrelation}

\subsection{Generating planetary model spectra} \label{Model generation}

We calculated synthetic model spectra for the cross-correlation and retrieval analysis using the radiative transfer code \mbox{petitRADTRANS} \citep{molliere2019PetitRADTRANSPythonRadiative}, and assumed an inverted $T$-$p$ profile characterised by the parametric model from \citet{guillot2010RadiativeEquilibriumIrradiated}. For the models used in the cross-correlation analysis, we chose an irradiation temperature $T_\mathrm{irr} = 2500$\,K, an internal temperature $T_\mathrm{int} = 200$\,K, an infrared opacity $\kappa_\mathrm{IR} = 10^{-1.7}$, and a ratio of visible to infrared opacity $\gamma = 10^{1.3}$. The resulting profile closely matches the one retrieved by \citet{yan2020TemperatureInversionAtomic} for WASP-189\,b.

We used the line lists noted in Table \ref{Table_SNR-Sysrem_iteration} to compute individual models for the following species that could be present under the expected conditions in WASP-189\,b's atmosphere: CO, CO$_2$, Fe, H$_2$O, FeH, NH$_3$, H$_2$S, HCN, and OH. We set the abundance of each species to a fixed value of $\log _{10} \mathrm{VMR} = -3$, and included continuum absorption of H- and collision-induced absorption (CIA) of H$_2$-H$_2$ and H$_2$-He \citep[and the references therein]{borysow2002CollisioninducedAbsorptionCoefficients, richard2012NewSectionHITRAN}. The resulting spectra were then converted into planet-to-star flux ratios, convolved to the previously determined resolution of CRIRES$^+$ (see Sect. \ref{Section_Data}) using a Gaussian instrumental profile, and continuum normalised.

\subsection{Cross-correlation of data and model}
We performed the computation of the weighted cross-correlation function (CCF) from the residual spectra and the model as described by \citet{cont2022AtmosphericCharacterizationUltrahot}:

\begin{align}
    \mathrm{CCF}(v, t) = \sum _{i = 0}^N \frac{R_i(t) \cdot M_i(v)}{\sigma_{i}(t)^2}\,,
\end{align}
where $t$ represents the time index, $i$ the pixel index, $R$ are the residual spectra with uncertainties $\sigma$, and $M$ the model spectrum, which was Doppler shifted with velocities $v$ ranging from -500\,km\,s$^{-1}$ to +500\,km\,s$^{-1}$. This resulted in a two-dimensional CCF map for each spectral segment. We combined the information from individual segments by computing the mean of these CCF maps, and merged the CCFs of both nodding positions.

We subsequently derived a $K_\mathrm{p}$-$v_\mathrm{offset}$ map from the CCFs by exploring a range of orbital semi-amplitude values ($K_p$) and shifting the CCF into the corresponding planetary frames based on the Doppler velocities:

\begin{align}
v_p = K_p \sin{2 \pi \phi} + v_\mathrm{sys} - v_\mathrm{bary} + v_\mathrm{offset}\,. \label{Eq_velocity}
\end{align}
Here, $\phi$ denotes the orbital phase, $v_\mathrm{sys}$ and $v_\mathrm{bary}$ are the systemic and barycentric velocities, and $v_\mathrm{offset}$ accounts for a potential deviation from the planetary rest frame. The shifted CCFs were then collapsed into a one-dimensional vector by averaging along the time axis. Stacking these vectors for every trial $K_p$ value resulted in a two-dimensional $K_\mathrm{p}$-$v_\mathrm{offset}$ map, which was converted into units of S/N by dividing by the standard deviation of the map in regions far away from the expected signal position ($|v| > 50$\,km\,s$^{-1}$).

\subsection{Selecting the number of \texttt{SYSREM} iterations} \label{Sysrem iterations}

When stellar and telluric lines are removed from the data using \texttt{SYSREM}, the number of iterations influences the resulting $K_\mathrm{p}$-$v_\mathrm{offset}$ map. Insufficient iterations leave strong residuals that impede the detection of planetary spectral signatures, while an excessive number of iterations risks removing the planetary signal. To objectively determine the optimal number of iterations, we implemented the method proposed by \hbox{\citet[see their Sect. 3.8]{cheverall2023RobustnessMeasuresMolecular}}. 

\begin{figure}
\centering
\includegraphics[width=\hsize]{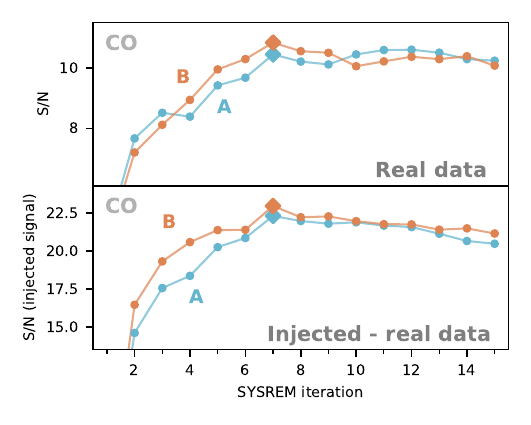}
  \caption{S/N detection strength as a function of \texttt{SYSREM} iterations for the signal of CO, at nodding positions A (blue colour) and B (orange colour). The top panel shows the signal strength for the real data, and the bottom panel shows the strength of an injected signal using the differential $\Delta$CCF according to the method described in Sect. \ref{Sysrem iterations}. The diamond shapes indicate the \texttt{SYSREM} iteration with the strongest detection of the injected signal.
          }
     \label{Fig_SNR-Sysrem_iteration}
\end{figure}

We first applied \texttt{SYSREM} to the normalised data, generating the cross-correlation function CCF$_\mathrm{obs}$. Then we repeated this process with an artificial signal injected at the expected Doppler velocity to produce CCF$_\mathrm{inj}$, from which we derived the differential cross-correlation function $\Delta$CCF = CCF$_\mathrm{inj}$ - CCF$_\mathrm{obs}$ and calculated the $K_\mathrm{p}$-$v_\mathrm{offset}$ map and S/N for each iteration. We chose the \texttt{SYSREM} iteration with the highest S/N in $\Delta$CCF and applied the same number of iterations to the non-injected data. Figure \ref{Fig_SNR-Sysrem_iteration} shows the S/N after each iteration for the example of CO, and the results of this process for all species are summarised in Table \ref{Table_SNR-Sysrem_iteration}.

\subsection{Cross-correlation signals}

\begin{table}[ht]\renewcommand{\arraystretch}{1.5}
    \caption[]{Summary of the cross-correlation results.}
    \label{Table_SNR-Sysrem_iteration}
    \begin{tabular}{lcccl}
        \hline \hline
        Species &
        \multicolumn{1}{p{0.5cm}}{\centering Iter. \\ A } &
        \multicolumn{1}{p{0.5cm}}{\centering Iter. \\ B } &
        \multicolumn{1}{p{1cm}}{\centering S/N \\ A + B } &
        Line list
        \\
        \hline

        CO & 7 & 7 & 14.2 & \citet{rothman2010HITEMPHightemperatureMolecular}\\
        Fe & 7 & 8 & 5.8 & \citet{kurucz2018ASPC..515...47K}\\
        % \hline
        FeH & 8 & 7 & 3.7 & \citet{wende2010AA...523A..58W}\\
        NH$_3$ & 8 & 6 & 3.1 & \citet{yurchenko2011VariationallyComputedLine}\\
        H$_2$O & 7 & 6 & 2.6 & \citet{polansky2018MNRAS.480.2597P}\\
        CO$_2$ & 7 & 6 & 2.2 & \citet{rothman2010HITEMPHightemperatureMolecular}\\
        OH & 7 & 7 & 2.0 & \citet{rothman2010HITEMPHightemperatureMolecular}\\
        HCN & 7 & 10 & 1.8 & \citet{harris2006ImprovedHCNHNC} \& \\
        & & & & \citet{barber2014ExoMolLineLists}\\ 
        H$_2$S & 7 & 9 & 1.8 & \citet{rothman2013JQSRT.130....4R}\\
        \hline
    \end{tabular}
    \tablefoot{For each species, the selected iteration for each nodding position and the combined S/N in the $K_\mathrm{p}$-$v_\mathrm{offset}$ map is given, as well as the reference for the line list used to compute the model.}
\end{table}

\begin{figure*}
  \includegraphics[width=\textwidth]{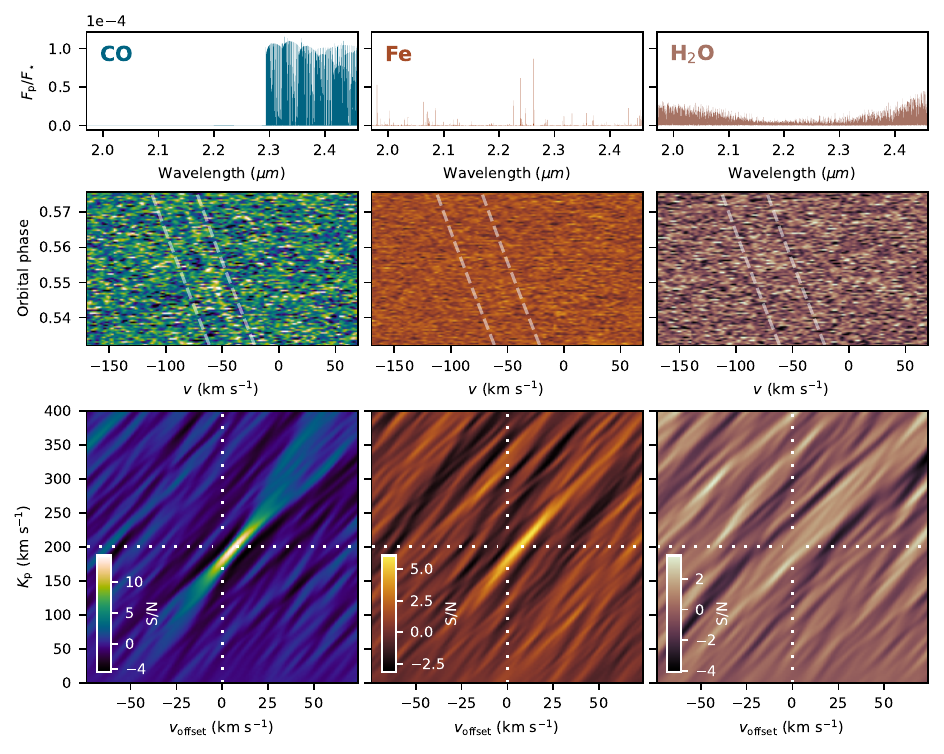}
  \caption{Model spectra (top), cross-correlation functions (CCFs, middle)), and $K_\mathrm{p}$-$v_\mathrm{offset}$ maps (bottom) of CO, Fe and H$_2$O. The dashed lines in the middle panels indicate the expected course of the planetary trail (which is located between the two lines). The signals of CO and Fe are visible in the CCFs and $K_\mathrm{p}$-$v_\mathrm{offset}$ maps, while H$_2$O was not detectable.}
  \label{Models+Detmaps}
\end{figure*}

We detected a cross-correlation signal for CO (S/N = 14.2) and Fe (S/N = 5.8), while the remaining species were not detectable. While we chose the number of \texttt{SYSREM} iterations using the method detailed above, the signals are stable across a wider range of iterations. The model spectra and $K_\mathrm{p}$-$v_\mathrm{offset}$ maps for each individual species are shown in Fig. \ref{Models+Detmaps} and \ref{Models+Detmaps_Undetected}. For both detected species we find an offset from the expected planetary rest frame velocity ($v_\mathrm{offset} = 6$\,km\,s$^{-1}$ for CO and $v_\mathrm{offset} = 7$\,km\,s$^{-1}$ for Fe), while the $K_\mathrm{p}$ agrees with the theoretical value within the error bars. 
This finding is in agreement with the red-shift of $4.5 ^{+2.2}_{-3.0}$\,km\,s$^{-1}$ measured in previous emission observations of CO \citep{yan2022DetectionCOEmission}. As detailed in their work, this could be evidence for global day-to-night winds, which have also been proposed by \citet{prinoth2021TitaniumOxideChemical} to explain a blue-shifted transmission signal.

\section{Atmospheric retrieval} \label{Sect_Retrieval}

\subsection{Treatment of the models} \label{Sect: Treatment of models}

We performed a retrieval analysis to quantitatively constrain the physical parameters of the planetary atmosphere. To this end, a Markov chain Monte Carlo (MCMC) algorithm sampled a large number of models with varying parameters, which were then compared to the residual spectra. These synthetic model spectra were generated in a similar way as described in Sect. \ref{Model generation}, and included all of the species that we probed for in the cross-correlation analysis. However, we did not assume isobaric abundances for the retrieval and instead determined pressure-dependent abundances based on equilibrium chemistry calculations using \texttt{fastChem} \citep{stock2018MNRAS.479..865S}. With this approach, the entire chemistry is determined by the atmospheric carbon-to-oxygen (C/O) ratio and metallicity [M/H], and the effect of thermal dissociation of molecules is included. We assumed that the bulk metallicity corresponds to the abundance of most atmospheric elements ([M/H] = [Fe/H] = [O/H] = [N/H] and so on), while the carbon abundance is set by the C/O ratio and oxygen abundance. As demonstrated by \citet{brogi2023AJ....165...91B}, isobaric abundances are an inadequate description for the extreme environments found in UHJ atmospheres and lead to conclusions inconsistent with the more appropriate equilibrium chemistry approach. In this work, we did not include disequilibrium effects such as photoionisation, which could influence the atmospheric abundance profiles of the highly irradiated UHJs. For the $T$-$p$ profile, we fixed $T_\mathrm{int} = 200$\,K as the observed spectra are not sensitive to the deeper atmospheric regions that are affected by this parameter.

The models were transformed into the same three-dimensional format as the residual spectra (segment $\times$ time $\times$ wavelength) and shifted into a planetary rest frame determined by the free parameters $K_\mathrm{p}$ and $v_\mathrm{offset}$ in the retrieval process. Additionally, we accounted for smearing of the signal due to the planet's change in velocity during each exposure. We calculated radial velocities at the start and end of each exposure (a difference of $\sim$3\,km\,s$^{-1}$), shifted the model to ten evenly spaced planetary rest frames between these velocities, and computed the smeared model as the mean over the ten sub-exposures.

To address rotational broadening effects, we computed a rotational profile from the equatorial rotation velocity $v_\mathrm{eq}$, following the method proposed by \citet{yan2023CRIRESDetectionCO} using Eq. (3) in \citet{diaz2011AccurateStellarRotational}. We assumed a linear limb darkening law with a coefficient $\epsilon = 1$ and an inclination angle of $\sin{i} \approx 1$, given the expected tidal locking of the planet. We convolved the model with this profile in logarithmic wavelength space, and finally brought the spectra to the resolution of CRIRES$^+$ using a Gaussian instrumental profile.

The application of \texttt{SYSREM} for the removal of stellar and telluric lines introduces distortions in the planetary signal. This effect should be accounted for to accurately match the models to the observed spectrum. While it is possible to directly apply \texttt{SYSREM} to the models to mimic the distortions, this would be computationally expensive and we instead applied an alternative model filtering following \citet{gibson2022RelativeAbundanceConstraints}. This approach involves preserving the individual vectors comprising each \texttt{SYSREM} model, and subsequently calculating a filter matrix from these results and the data uncertainties. Such a filter can be pre-computed and then applied to each individual model during the retrieval process. We analyse the performance of this filter and its ability to preserve subtle differences in the line profile in App. \ref{Appendix_Filter}.

\subsection{Likelihood function}
We calculated the logarithmic likelihood function \citep{hogg2010DataAnalysisRecipes, yan2020TemperatureInversionAtomic} to compare each model spectrum to the data:

\begin{align}
\ln(L) = - \frac{1}{2}\sum_{i,j}\left(\frac{(R_{i,j} - M_{i,j})^2}{(\beta \sigma_{i,j})^2} + \ln(2\pi(\beta \sigma_{i,j})^2)\right)\,,
\end{align}
where $R_{i,j}$ denotes the residual spectrum at pixel $i$ and time $j$, $M_{i,j}$ represents the 2D matrix of a filtered model spectrum, $\sigma_{i,j}$ are the uncertainties of the residuals, and $\beta$ is a scaling factor for the $\sigma_{i,j}$. The parameter space was sampled using an MCMC algorithm as implemented in $\texttt{emcee}$ \citep{foreman-mackey2013EmceeMCMCHammer}, with 32 walkers and 15\,000 steps each. We discarded the first 5000 steps as burn-in, and assessed the convergence by checking the autocorrelation lengths as well as dividing the chains into multiple sets and ensuring that all sub-sets result in similar posteriors. We chose uniform priors for all free parameters, which we summarised in Table \ref{Table_Priors}.

\subsection{Retrieval results}

\begin{figure*}
  \includegraphics[width=\textwidth]{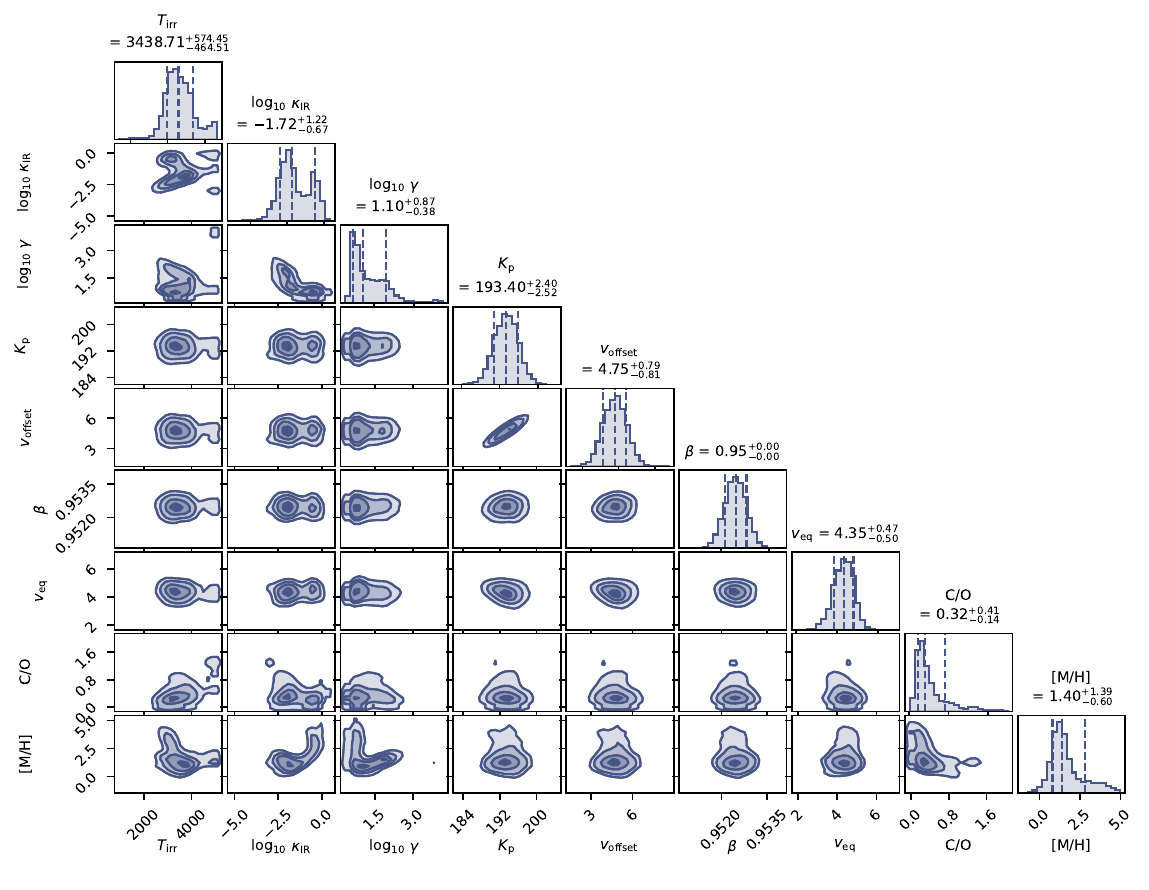}
  \caption{Posterior distributions from the retrieval of the atmospheric and orbital parameters. The dashed lines indicate the mean value and the 16th and 84th percentile. The included parameters consists of the three parameters defining the $T$-$p$ profile ($T_\mathrm{irr}$, $\kappa_\mathrm{IR}$, $\gamma$), the orbital semi-amplitude of the planet ($K_\mathrm{p}$), the offset from the expected planet rest frame ($v_\mathrm{offset}$), a scaling factor to the uncertainties ($\beta$), the equatorial rotation velocity ($v_\mathrm{eq}$) and two parameters determining the equilibrium chemistry (C/O ratio and metallicity).}
  \label{Fig_Cornerplot_AllSpecies}
\end{figure*}

\begin{table}[ht]\renewcommand{\arraystretch}{1.5}
 \caption[]{Priors and retrieved values of the initial atmospheric retrieval.}\label{Table_Priors}
\begin{tabular}{llll}
 \hline \hline
  Parameter &
  Prior &
  Retrieved value &
  Unit

 \\ \hline
% $T_\mathrm{int}$   & [100, 3000] & 1100$^{+600}_{-700}$ &  K  \\
$T_\mathrm{irr}$   & [300, 5000] & 3440 $^{+570}_{-460}$ &  K \\
$\log_{10}$ $\kappa_\mathrm{IR}$   & [-15, 4] & $-1.7^{+1.2}_{-0.7}$ &  $\log_{10}$ cm$^2$\,s$^{-1}$  \\
$\log_{10}$ $\gamma$  & [-2, 2] & $1.1^{+0.9}_{-0.4}$  &   - \\
$K_p$   & [170, 230] & $193.4^{+2.4}_{-2.5}$  &  km\,s$^{-1}$  \\
$v_\mathrm{offset}$   & [-20, 20] & $4.7\pm 0.8$  &  km\,s$^{-1}$  \\
$v_\mathrm{eq}$   & [0.1, 20] & $4.35^{+0.47}_{-0.50}$  &  km\,s$^{-1}$  \\
$\beta$   & [0.2, 5] & $0.95\pm0.0003 $  &   - \\
C/O  & [0, 2] & $0.32^{+0.41}_{-0.14}$  &  -  \\\relax
[M/H]   & [-5, 5] & $1.40^{+1.39}_{-0.60}$  &   dex \\
\hline
\end{tabular}
\end{table}

The posterior distributions of the free parameters are shown in Fig. \ref{Fig_Cornerplot_AllSpecies} and summarised in Table \ref{Table_Priors}. The scaling factor $\beta$, which is applied to the data uncertainties, is close to unity, indicating that the uncertainties determined during the data reduction are accurate. The remaining free parameters determining the $T$-$p$ profile, chemistry, and velocities are discussed in the following.

\subsubsection{Temperature-pressure profile}

\begin{figure*}
\centering
\includegraphics[width=\hsize]{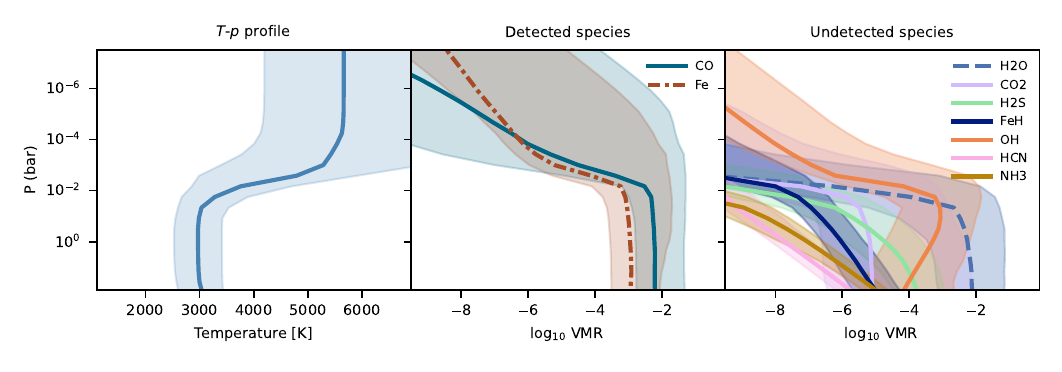}
  \caption{Retrieved temperature and abundance profiles. Left panel: Retrieved $T$-$p$ profile calculated from 10\,000 random samples. The shaded region indicates the 1$\sigma$ uncertainty interval. Middle and right panel: Retrieved abundance profiles of the individual species as determined from the C/O and [M/H] values and the $T$-$p$ profiles of 1000 random samples. The two species in the middle panel have been detected in the cross-correlation analysis, while the species in the right panel were undetectable.}
     \label{Fig_Tp_Abundances}
\end{figure*}
Observations of the dayside emission are mostly sensitive to the properties of the thermal inversion in the atmosphere, because the emission lines originate from that region. We therefore expect to constrain the pressure range at which the inversion is located and the temperature difference between the deeper, colder  layer and the upper, warmer layer. The retrieved values of the three free parameters defining the $T$-$p$ profile ($T_\mathrm{irr}$, $\kappa_\mathrm{IR}$, and $\gamma$) are correlated with each other. This is due to the fact that various combinations of these parameters can lead to similar temperature inversions.

Larger values of the visible to infrared opacity $\gamma$ are correlated with lower infrared opacities $\kappa_\mathrm{IR}$ and reduced irradiation temperatures $T_\mathrm{irr}$. As $\gamma$ increases, the inversion layer shifts to higher altitudes and the temperature difference increases. This effect can be mitigated by decreasing $\kappa_\mathrm{IR}$, which results in a lower inversion altitude.  Additionally, reducing $T_\mathrm{irr}$ lowers the temperature across the entire upper atmosphere down to pressures of $\sim$1\,bar. These adjustments thus achieve a $T$-$p$ profile closely resembling that associated with lower $\gamma$ values.

The average retrieved $T$-$p$ profile, derived from 10\,000 randomly drawn samples, is shown in Fig. \ref{Fig_Tp_Abundances}. The profile is well constrained between $10^{-1}$\,bar and $10^{-3}$\,bar, which is the region probed by our high-resolution observation.  However, the profile cannot be constrained at lower pressures because the observed spectra remain insensitive to the temperatures at these altitudes. The same is true for the temperature of the deep atmospheric layers, which is artificially constrained in the retrieval because we fixed $T_\mathrm{int}$ to a constant value.

\subsubsection{Chemistry}
The chemical composition in our models is determined by the atmospheric C/O ratio and metallicity, and we retrieved C/O = $0.32^{+0.41}_{-0.14}$ and [M/H] = $1.40^{+1.39}_{-0.60}$ by assuming equilibrium chemistry. However, the posterior distribution of C/O showed a long tail towards higher values and is fully consistent with the solar C/O ratio. A major reason for the inability to tightly constrain the C/O ratio stems from the non-detection of other oxygen-bearing species besides CO, such as H$_2$O or OH. The information about the abundances of CO and Fe alone was not sufficient to precisely determine the C/O ratio, especially due to the effect the $T$-$p$ profile has on the emission spectrum. The high C/O values were generally associated with smaller $\kappa_\mathrm{IR}$ and elevated $T_\mathrm{irr}$, resulting in models with similar strengths of CO and Fe emission lines. In our chemistry model, the abundance of oxygen is linked to that of iron, and therefore our retrieved C/O ratio is only accurate if we assume that [O/H] = [Fe/H].

Although we consistently retrieved super-solar metallicities, the precise value  remained uncertain due to a degeneracy with C/O. Higher [M/H] correlated with lower C/O, as these combinations yield similar CO abundances. While the abundance measurement of Fe should theoretically help in determining [M/H], it appears insufficient to break the degeneracy in practice. Due to the large uncertainties, the metallicity is consistent with the solar value within 3$\sigma$.

If the metallicity of WASP-189\,b is truly higher than the solar metallicity of the host star \citep{anderson2018WASP189bUltrahotJupiter}, this would imply a significant enrichment of heavy elements. In combination with a sub-solar to solar C/O, this would hint at an accretion of icy solids outside of the H$_2$O snowline \citep{Oeberg2011ApJ...743L..16O}. An alternative pathway to such a combination of C/O and metallicity is an intensive erosion of the planets core and a distribution of the material across the gaseous envelope \citep{Madhusudhan2017MNRAS.469.4102M}.

We derived the volume mixing ratios (VMRs) of all included species from the retrieved chemical and $T$-$p$ parameters by averaging over 1000 random samples (see Fig. \ref{Fig_Tp_Abundances}). The $\log$(VMR) of the detected species CO and Fe was constrained within uncertainties of $\sim$ 1\,dex in the altitude range of the inversion layer ($10^{-1}$ -- $10^{-3}$\,bar). However, we could not reliably determine their abundances above these pressure levels, because the observed spectrum is insensitive to changes in these regions. \citet{Gandhi2023AJ....165..242G} retrieved a Fe abundance of $\log$(VMR$_\mathrm{Fe}$) = $-4.94^{+0.24}_{-0.26}$ from high-resolution transit observations, which is consistent with our abundance profile in the upper atmosphere. It is a reasonable assumption that the signal in their work originates from these higher regions, as they additionally retrieve an opacity deck that blocks contributions from below 10$^{-3}$\,bar. This could also be the reason for the low metallicity found in their work, which is based on the Fe abundance and falls significantly below our retrieved metallicity value.

Species without detectable cross-correlation signals were included in the retrieval to discourage combinations of C/O and [M/H] that result in abundances which would be detectable with our observation. According to the retrieved equilibrium chemistry, we expect significant amounts of H$_2$O and OH in the lower atmosphere, with only traces of  FeH, NH$_3$, CO$_2$, HCN and H$_2$S. None of these species persist in the upper atmosphere due to thermal dissociation. The retrieved abundance profiles are consistent with the non-detections in cross-correlation, as confirmed by injection tests.

\subsubsection{Velocities}
The retrieval resulted in a planetary radial velocity semi-amplitude of $K_p = 193.4^{+2.4}_{-2.5}$\,km\,s$^{-1}$, falling within the 2$\sigma$ range of the value derived from literature orbital parameters ($K_{p,\mathrm{lit}} = 201\,\pm\,4$\,km\,s$^{-1}$). Additionally, we found a significant velocity offset from the expected planetary rest frame of $v_\mathrm{offset} = 4.7\,\pm\,0.8$\,km\,s$^{-1}$. Both results agree with the signal position found in the cross-correlation analysis, and with the findings of previous high-resolution emission observations of this planet \citep{yan2020TemperatureInversionAtomic, yan2022DetectionCOEmission}.

We assumed that the planetary atmosphere is rigidly rotating and retrieved an equatorial velocity of $v_\mathrm{eq} = 4.35^{+0.47}_{-0.50}$\,km\,s$^{-1}$. Including a contribution of 3.0\,km\,s$^{-1}$ from tidally locked rotation, this indicates an excess speed of $1.35^{+0.47}_{-0.50}$\,km\,s$^{-1}$ for the planetary surface rotation pattern. These velocities offer initial insights into the wind dynamics on the dayside of WASP-189\,b: a red-shift of the observed signal can be caused by day-to-night winds, while an increased rotational broadening suggests the presence of zonal equatorial jets, which are expected to form on UHJs \citep[e.g.][]{lee2022MantisNetworkII}. To investigate which of these processes is more prevalent, we performed an additional retrieval that includes a more sophisticated modelling of the effects of winds on the line profile in the following section.

\section{Retrieval of the wind pattern} \label{Sect_WindRetrieval}
\subsection{Calculating velocity profiles}

\begin{figure*}[h]
\centering
\includegraphics[width=\hsize]{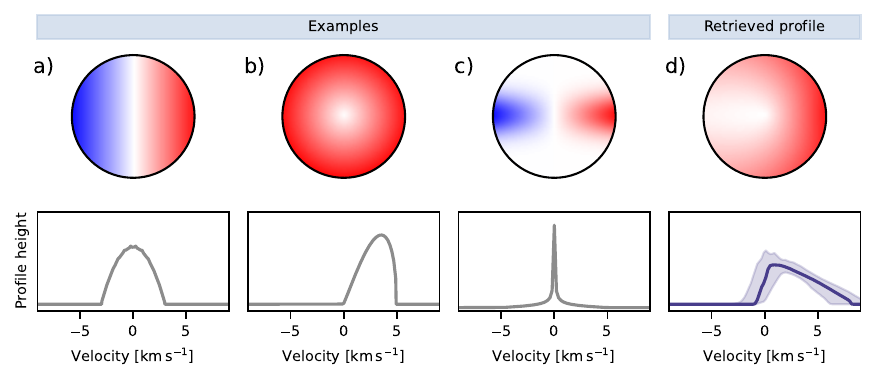}
  \caption{Examples of different radial velocity distributions on the planetary disk and resulting velocity profiles, and the result from the retrieval. The example profiles assume a limb darkening coefficient of $\epsilon = 1$. a) Tidally locked rotation (according to WASP-189\,b's orbital period). b) Day-to-night wind of $v_\mathrm{day-night}=5$\,km\,s$^{-1}$. c)  Zonal jet with a speed of $v_\mathrm{jet}=5$\,km\,s$^{-1}$ and a width of $\sigma_\mathrm{jet}=0.3$\,R$_\mathrm{p}$. d) Velocity distributions according to the best-fitting parameters from the retrieval, and the average retrieved line profile calculated from 1000 random samples. The shaded region indicates the 1$\sigma$ uncertainty interval.}
     \label{Fig_Velocity_Profiles}
\end{figure*}

To retrieve the atmospheric wind pattern, we determined an empirical line profile by dividing the planetary disc into a grid of quadratic cells, with the planet's diameter corresponding to 2001 cells. Then we assigned a radial velocity and brightness to each cell and computed a velocity profile from these data. The velocities of different regions of the atmosphere are caused by a combination of multiple effects: 
\begin{itemize}
    \item[$\circ$] Tidal locking induces a uniform rotation of the planet with a rotation period equal to the orbital period. In the case of WASP-189\,b, this leads to a maximal velocity of 3.0\,km\,s$^{-1}$. The projected radial velocity grows linearly with the distance from the axis of rotation.
    \item[$\circ$] Day-to-night winds flow from the sub-stellar point to the nightside in all directions, leading to an overall red-shift. We assumed a constant wind speed $v_\textrm{day-night}$ across the entire visible hemisphere, and the resulting radial velocities depend solely on the angle to the line of sight. 
    \item[$\circ$] Zonal jets typically cause a super-rotation of the regions close to the equator. We assumed that the velocity decreases towards the poles following a Gaussian profile, and parametrised such a jet with its equatorial velocity $v_\mathrm{jet}$ and width $\sigma_\mathrm{jet}$ (describing the standard deviation of the Gaussian distribution in units of $R_\mathrm{p}$).
\end{itemize}
We did not directly include a hot spot in our simulation in order to reduce the computational complexity. Instead, we approximated its effect by modulating the brightness of each cell following a linear limb-darkening law with a coefficient $\epsilon$ as a free parameter. In this way, the central region has a larger contribution compared to the terminator at the outer edge of the planetary disk.

The resulting velocity profile was determined by calculating a histogram of the radial velocities while weighting each cell by its brightness. Examples for the velocity distributions and profiles for tidally locked rotation, a day-to-night wind, and a jet are shown in Fig. \ref{Fig_Velocity_Profiles} a), b), and c), respectively. To include this framework in the model spectra calculation, we replaced the rotational profile described in Sect. \ref{Sect: Treatment of models} with this more complicated velocity profile, and convolved it with the models in logarithmic wavelength space.

Compared to the previously conducted retrieval shown above, we have thus replaced the free parameter $v_\mathrm{eq}$ with $v_\mathrm{jet}$, $\sigma_\mathrm{jet}$, $v_\textrm{day-night}$, and $\epsilon$, and assumed the tidally locked rotation to be always present. All other parameters and the included species remained the same.

\subsection{Retrieved wind parameters}

\begin{figure*}
  \includegraphics[width=\textwidth]{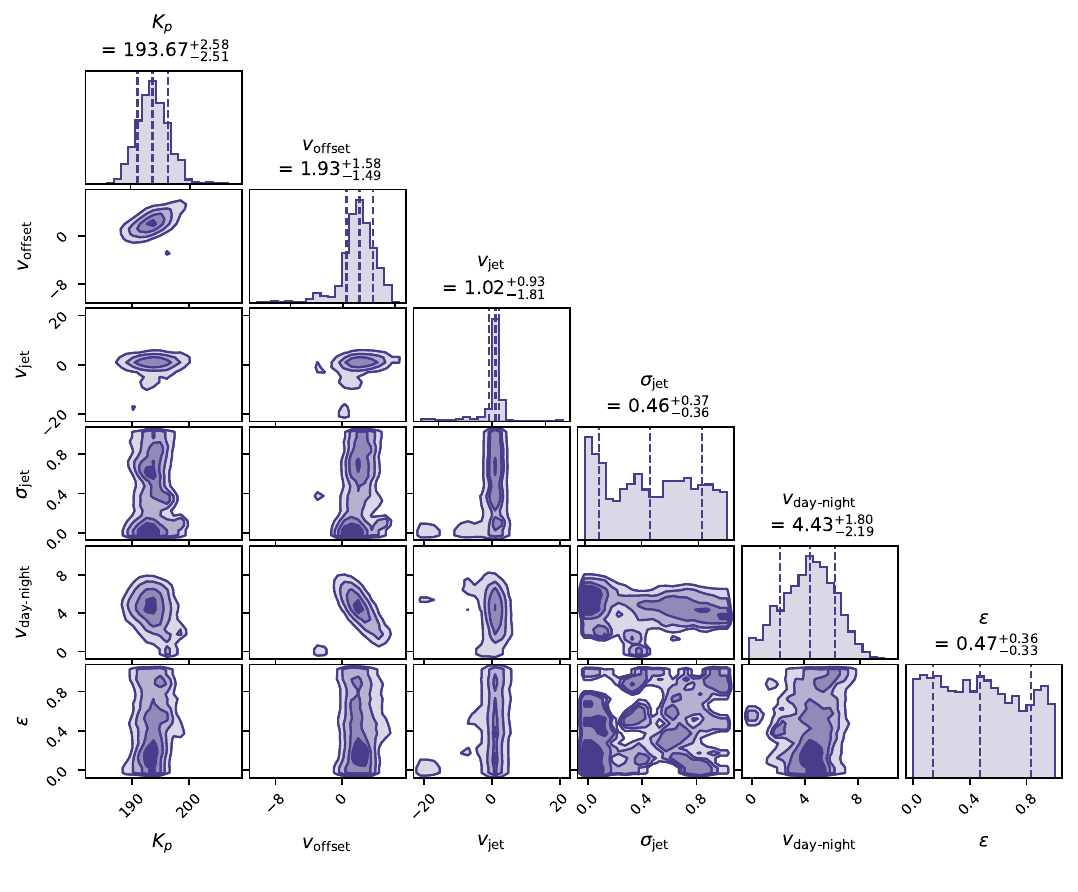}
  \caption{Posterior distributions from the retrieval of the atmospheric and orbital parameters. The dashed lines indicate the mean value and the 16th and 84th percentile. The posteriors of the jet width $\sigma_\mathrm{jet}$ and limb darkening coefficient $\epsilon$ fill the entire parameter space and the percentile values of these two parameters are thus not meaningful.}
  \label{Fig_Cornerplot_Windparameter}
\end{figure*}

We found that the observed line profile and velocity offset are best fitted by a model dominated by a fast day-to-night wind. The posterior distributions of the wind parameters are shown in Fig. \ref{Fig_Cornerplot_Windparameter}. The velocity of the day-to-night wind was retrieved as $v_\textrm{day-night} = 4.4^{+1.8}_{-2.2}$\,km\,s$^{-1}$, and the jet velocity as $v_\mathrm{jet} = 1.0^{+0.9}_{-1.8}$\,km\,s$^{-1}$, which is consistent with the over-rotation of the first retrieval. The width of the jet ($\sigma_\mathrm{jet}$) remained unconstrained, which is likely due to the low velocity of the jet, because the effect of the jet width on the line profile in such a scenario is minor, leading to weak constraints. 
We were not able to place constrains on the limb darkening coefficient $\epsilon$, likely due to its minute effect on the spectra which was not measurable in our data. The retrieved line profile as calculated from averaging over 1000 random samples is shown in Fig. \ref{Fig_Velocity_Profiles} d).

Compared to the retrieval without winds, the velocity offset was significantly reduced to $v_\mathrm{offset} = 1.9^{+1.6}_{-1.5}$\,km\,s$^{-1}$. This is because a day-to-night wind induces an overall red-shift of the planetary signal, which can account for some of the offset. However, the offset can not be completely explained by our retrieved wind scenario, because this would require wind speeds upwards of 7\,km\,s$^{-1}$, which would lead to an extensive broadening that is inconsistent with the observed broadening. Other effects which have not been included in our model could account for the remaining offset from the expected planetary rest frame. The increased flux from a hot spot close to the sub-stellar point might dominate the observed emission signal, and as the hot spot rotates away from the observer following the tidally locked rotation, an overall red-shift of the spectrum could be produced. We assumed that the full dayside is visible during the entire observation, while in reality the planet rotates so that more and more of the nightside comes into view. This effect is not very pronounced in our observation, which is close to the secondary eclipse and only covers a small section of the orbit (0.53 < $\phi$ < 0.57). According to our simulations, a hot spot alone can not result in the retrieved line profile, and a day-night wind is required to fit the extended red-shifted wing of the line profile. Nevertheless, the hot spot could introduce a small red-shift as the location of the nightside is associated with negative radial velocities in our viewing geometry.
Alternatively, the offset could be due to uncertainties in the ephemerides and system parameters. The propagated error of the orbital period corresponds to a velocity uncertainty of $\sim$1\,km\,s$^{-1}$, and an error of similar scale could be introduced by the uncertainties in the systemic velocity due to extensive stellar line broadening of the rapidly rotating host star and inaccuracies of the stellar models. Further, a slight eccentricity (0 < $e$ < 0.005) could cause an offset of the correct magnitude while still being consistent with the observed signal within the uncertainties. Additionally, the uncertainty of the planetary radius propagates to the tidally locked rotation velocity, which in turn affects the line broadening. A larger radius would result in stronger broadening from the rotation alone, which in turn would reduce the day-to-night wind speed still consistent with the observed broadening. As the radius uncertainties from the CHEOPS observations of \citet{lendl2020HotDaysideAsymmetric} are on the order of 1\%, we do not expect this effect to significantly influence our results. In contrast to the absolute offset, the relative offset between emission and transmission signal is not affected by these systematic effects. In the future, a joint retrieval of both types of observations could result in a consistent model that can explain all of the observed velocity shifts.

We note that the retrieved posteriors of the day-to-night wind speed are consistent with a wind-free scenario to within $\sim$2$\sigma$, and these large uncertainties are mainly caused by the degeneracy between wind speed and velocity offset. In a wind-free scenario, the entire offset of $\sim$5\,km\,s$^{-1}$ would have to be caused by other effects such as those mentioned above, while in the retrieved best-fitting scenario the remaining offset is reduced to 2\,km\,s$^{-1}$.

The full corner plot of all free parameters for this retrieval with a dedicated wind line profile is shown in Fig. \ref{Fig_Cornerplot_Windparameter_full}, and a summary of the priors and retrieved values can be found in Table \ref{Table_Priors_WindRetrieval}. We found that the $T$-$p$ profile parameters, C/O ratio, metallicity, and $K_p$ are consistent within 1\,$\sigma$ with the previous retrieval, which did not include a model of the wind pattern and only accounted for a general line broadening.

While the presence of both jets and day-to-night winds is predicted for atmospheres of hot Jupiters \citep{zhang2017ConstrainingHotJupiter}, the models of \citet{Showman2013ApJ...762...24S} showed that the equatorial jets can be inhibited due to radiative and frictional damping for planets under extreme stellar irradiation. In this case the day-to-night flows would dominate, which agrees with the best-fitting scenario from our retrieval. The retrieved weak equatorial jet also agrees with the conclusion of \citet{tan2019ApJ...886...26T}, who simulated the dependence of the jet strength on the equilibrium temperature and rotation period. They found that higher temperatures lead to weaker jets, because the dissociation of H$_2$ on the dayside and subsequent recombination on the nightside decreases the day-night temperature gradient. This effect could be one explanation for the apparent slower wind speeds on the ultra-hot WASP-189\,b compared to some cooler hot Jupiters, such as WASP-43\,b \citep{lesjak2023A&A...678A..23L} or WASP-166\,b \citep{Seidel2020A&A...641L...7S}. The 3D general circulation models of \citet{lee2022MantisNetworkII} predict a fast jet of 5\,km\,s$^{-1}$ in WASP-189\,b's atmosphere, which is at odds with our result. However, their work did not include the dissociation of H$_2$, which could lead to an overestimation of the day-night temperature gradient and consequently result in higher jet speeds. Additionally, their model was not able to fully reproduce the flux variation from the CHEOPS phase curve data. While they find that their model agrees with the results of \citet{prinoth2021TitaniumOxideChemical}, we note that in the transmission spectrum both day-to-night winds and equatorial jets can cause blue-shifts of the detected signals. In the case of a jet, the material at the hotter evening terminator moves towards the observer and contributes more to the signal than the material on the cooler morning terminator. When observing the dayside of a planet, equatorial jets mainly cause a line broadening while the introduced velocity shift is minute. Therefore it is difficult to explain the observed offset in the emission spectrum only with a jet.

The retrieved day-to-night wind speed is in general agreement with the observed blue-shifts in the transmission signal of $\sim$5\,km\,s$^{-1}$ from previous studies. \citet{prinoth2021TitaniumOxideChemical} found different offsets for each of the detected species, and in particular their value for Fe of $3.63 \pm 0.43$\,km\,s$^{-1}$ is slightly below our retrieved value but agrees within our uncertainties. We note that our wind properties are mostly determined based on the CO line shape, because the signal of CO was significantly larger than that of Fe.

While our model aims to determine the resulting line profiles of various scenarios of atmospheric dynamics, it is based on some simplifying assumptions that warrant consideration. We assumed uniform day-to-night wind speeds and a fixed viewing geometry, which excluded contributions from the nightside. Additionally, the model did not account for a sub-stellar hot spot, which could increase flux from a localised region of the atmosphere and enhance the ionisation of molecular species. To fully capture the complexities of the planet's atmospheric circulation, a future study might combine transmission and dayside observations and employ a more complex 3D modelling of the line profile which incorporates these effects.

\section{Summary} \label{Section_Conclusion}

We conducted a comprehensive analysis of the dayside atmosphere of WASP-189\,b using high-resolution spectroscopic observations with CRIRES$^+$.  The cross-correlation analysis confirmed the presence of CO and Fe with signals red-shifted by $\sim$6\,km\,s$^{-1}$, while a range of other species were not detectable. By performing atmospheric retrieval analyses, we quantitatively constrained the atmospheric and orbital parameters, and we incorporated a detailed modelling of the line profile due to various dynamical effects in the retrieval framework.

Assuming equilibrium chemistry, we found a C/O ratio consistent with the solar value, and a stellar to super-stellar metallicity, consistent with models of late-stage ice accretion or core erosion.
The effects of winds on the observed spectra were modelled based on the radial velocity distributions of different wind patterns on a 2D disk representing the planetary dayside.
The retrieved line profile fits best with an atmosphere primarily influenced by a fast day-to-night wind, inducing a notable red-shift in the observed spectra. In contrast, the retrieved equatorial jet velocity was significantly lower and the width of the jet remained unconstrained. While some theoretical works predict slower wind speeds in ultra-hot atmospheres, other general circulation models of WASP-189\,b result in significantly faster jet speeds that are not consistent with the retrieved line profile. Not the entire offset from the expected planet rest frame can be explained by day-to-night winds, because a further increase of the wind speed not only shifts the signal position but also causes a further broadening, which is not consistent with the observed data. Because of this effect, it is generally difficult to explain large velocity offsets in any dayside spectra purely with day-to-night winds. Possible reasons for the remaining offset in our observation include an inaccurate system velocity and the effect of a sub-stellar hot spot.

Further observations covering a wider range of orbital phases before and after the eclipse will be beneficial to further constrain the dynamical properties. The combination with transmission observations will likely prove useful in explaining the remaining velocity offset and providing a more comprehensive understanding of the atmospheric circulation and thermal structure of WASP-189\,b.

\begin{acknowledgements}
CRIRES$^+$ is an ESO upgrade project carried out by Thüringer Landessternwarte Tautenburg, Georg-August Universität Göttingen, and Uppsala University. The project is funded by the Federal Ministry of Education and Research (Germany) through Grants 05A11MG3, 05A14MG4, 05A17MG2 and the Knut and Alice Wallenberg Foundation. Based on observations collected at the European Organisation for Astronomical Research in the Southern Hemisphere under ESO programme 109.23HN.002. F.L. acknowledges the support by the Deutsche Forschungsgemeinschaft (DFG, German Research Foundation) – Project number 314665159. D.C. is supported by the LMU-Munich Fraunhofer-Schwarzschild Fellowship and by the Deutsche Forschungsgemeinschaft (DFG, German Research Foundation) under Germany´s Excellence Strategy – EXC 2094 – 390783311. F.Y. acknowledges the support by the National Natural Science Foundation of China (grant No. 42375118). L.B.-Ch., A.D.R., and N.P. acknowledge support by the Knut and Alice Wallenberg Foundation (grant 2018.0192). S.C. acknowledges the support of the DFG priority program SPP 1992 “Exploring the Diversity of Extrasolar Planets" (CZ 222/5-1). M.R. acknowledges the support by the DFG priority program SPP 1992 “Exploring the Diversity of Extrasolar Planets” (DFG PR 36 24602/41). D.S. acknowledges financial support from the project PID2021-126365NB-C21(MCI/AEI/FEDER, UE) and from the Severo Ochoa grant CEX2021-001131-S funded by MCIN/AEI/ 10.13039/501100011033.
\end{acknowledgements}

% WARNING
%-------------------------------------------------------------------
% Please note that we have included the references to the file aa.dem in
% order to compile it, but we ask you to:
%
% - use BibTeX with the regular commands:
%   \bibliographystyle{aa} % style aa.bst
%   \bibliography{Yourfile} % your references Yourfile.bib
%
% - join the .bib files when you upload your source files
%-------------------------------------------------------------------
\bibliographystyle{aa} % style aa.bst
\bibliography{WASP-189b_Bibliography}

\newpage
\appendix

\onecolumn
\section{Additional $K_p$\,-\,$v_\mathrm{sys}$ maps}

\begin{center}
   \includegraphics[width=\textwidth]{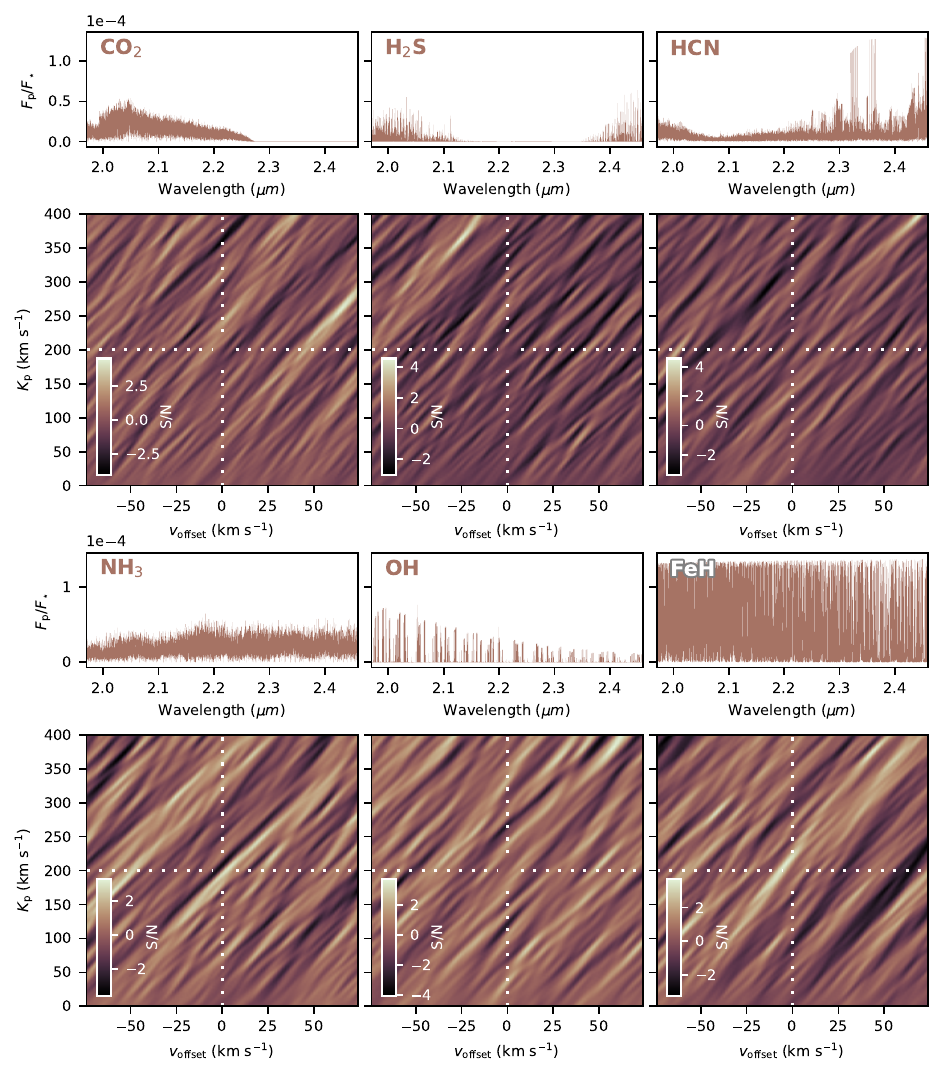}
  \captionof{figure}{Model spectra and $K_\mathrm{p}$-$v_\mathrm{offset}$ maps of the six undetected species CO$_2$, H$_2$S, HCN, NH$_3$, OH, and FeH.}
  \label{Models+Detmaps_Undetected}
\end{center}

\onecolumn
\section{Evaluating the effect of the model filter} \label{Appendix_Filter}

\FloatBarrier
\begin{figure*}[h!]
        \centering
        \includegraphics[width=\textwidth]{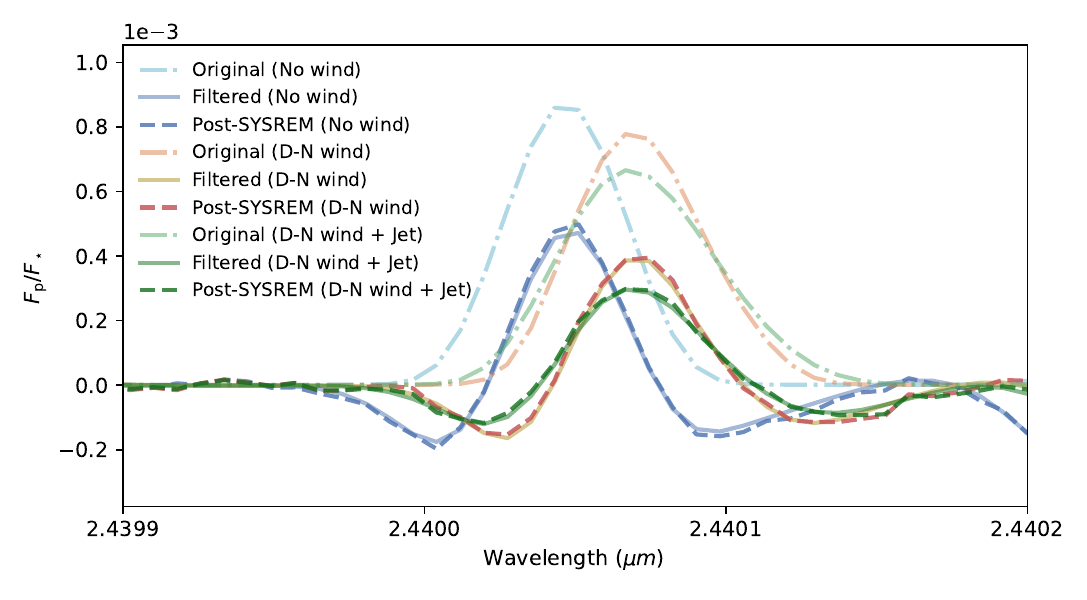}
        \caption{Comparison of spectral line shapes of synthetic CO models for three scenarios: without any atmospheric dynamics (blue), with a day-to-night wind of 5\,km\,s$^{-1}$ (red), and with a day-to-night wind of 5\,km\,s$^{-1}$ and a jet of 3\,km\,s$^{-1}$ (green). The unprocessed models (dash-dotted lines) were injected into the observed data, to which subsequently \texttt{SYSREM} was applied for 4 iterations. The ratio of the resulting residuals to the residuals of non-injected data show the distortions of the planetary signal introduced by \texttt{SYSREM} (dashed lines). These distortions can be reproduced by applying a filter directly to the original models, resulting in the filtered models (solid lines) which closely resemble the injected models in all three scenarios.}
     \label{Fig_GibsonFilter_LineProfiles}
\end{figure*}
\FloatBarrier
The application of \texttt{SYSREM} can alter the planetary spectrum in the observed data. To account for this effect in our model spectra, we apply a filter based on the approach of \citet{gibson2022RelativeAbundanceConstraints}, as outlined in Sect. \ref{Sect: Treatment of models}. This enables a more accurate comparison between the observed and modelled spectra during the retrieval process. Since our analysis revolves around the line shapes of different planetary features, it is crucial that the filtering process preserves these shapes without introducing major distortions. In the following, we evaluate the filter's performance by comparing the filtered models to the actual effect of \texttt{SYSREM} on an injected model spectrum.

To this end, we injected a model spectrum into the observed data, applied \texttt{SYSREM}, and divided the result by the non-injected residuals. The ratio of injected to non-injected residuals represents the planetary signal post-SYSREM, which showed clear distortions compared to the original model. Figure \ref{Fig_GibsonFilter_LineProfiles} illustrates the comparison between a single CO emission line from the original model, the post-SYSREM model, and the result after applying the filter. Shown are three different scenarios: without any atmospheric dynamics, with a day-to-night wind of 5\,km\,s$^{-1}$, and with a day-to-night wind as well as a jet of 3\,km\,s$^{-1}$. Although this example used four iterations of \texttt{SYSREM}, a similar behaviour is observed across other iterations.

SYSREM primarily reduces the amplitude of spectral lines and introduces troughs on either side of the lines. The filtered model accurately replicates these effects, with only minor deviations from the injected model. Importantly, the differences in line shape between the models---such as the broader line width and extended redshifted wing in the wind-included models---are preserved. Although a wind speed of 5\,km\,s$^{-1}$ was used in this demonstration, the line profile differences are retained also at lower wind speeds. Thus, we conclude that applying this filter to the models before comparing them with \texttt{SYSREM}-processed data enables reliable determination of wind speeds based on the spectral line shapes.

% \onecolumn
\section{Full corner plot and parameters of the wind pattern retrieval}
\begin{center}
   \includegraphics[width=\textwidth]{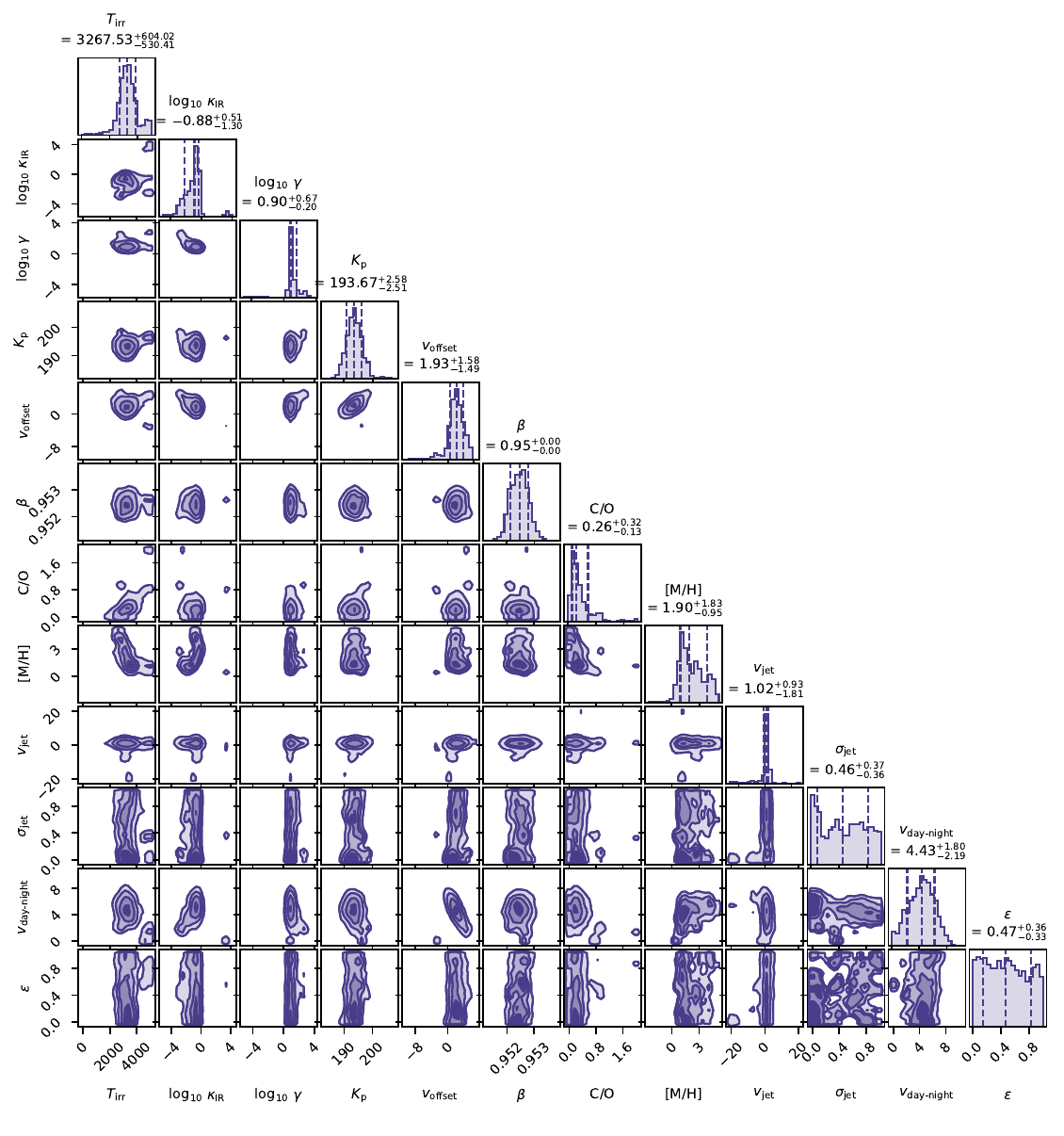}
  \captionof{figure}{Full corner plot of the posterior distributions from the retrieval with a dedicated wind line profile. The dashed lines indicate the mean value and the 16th and 84th percentile. The posteriors of the jet width $\sigma_\mathrm{jet}$ and limb darkening coefficient $\epsilon$ fill the entire parameter space and the percentile values of these two parameters are thus not meaningful.}
  \label{Fig_Cornerplot_Windparameter_full}
\end{center}

\begin{table}[ht]\renewcommand{\arraystretch}{1.5}
\centering
 \caption[]{Priors and retrieved values of the wind pattern retrieval.}\label{Table_Priors_WindRetrieval}
\begin{tabular}{llll}
 \hline \hline
  Parameter &
  Prior &
  Retrieved value &
  Unit

 \\ \hline
% $T_\mathrm{int}$   & [100, 3000] & 1100$^{+600}_{-700}$ &  K  \\
$T_\mathrm{irr}$   & [300, 5000] & 3270 $^{+600}_{-530}$ &  K \\
$\log_{10}$ $\kappa_\mathrm{IR}$   & [-5, 5] & $-0.9^{+0.5}_{-1.3}$ &  $\log_{10}$ cm$^2$\,s$^{-1}$  \\
$\log_{10}$ $\gamma$  & [-5, 5] & $0.9^{+0.7}_{-0.2}$  &   - \\
$K_p$   & [170, 230] & $193.7^{+2.6}_{-2.5}$  &  km\,s$^{-1}$  \\
$v_\mathrm{offset}$   & [-20, 20] & $1.9^{+1.6}_{-1.5}$  &  km\,s$^{-1}$  \\
$\beta$   & [0.2, 5] & $0.95\pm0.0003 $  &   - \\
C/O  & [0, 2] & $0.26^{+0.32}_{-0.13}$  &  -  \\\relax
[M/H]   & [-5, 5] & $1.90^{+1.83}_{-0.95}$  &   dex \\
$v_\mathrm{jet}$   & [-20, 20] & $1.02^{+0.93}_{-1.81}$  &  km\,s$^{-1}$  \\
$\sigma_\mathrm{jet}$   & [0, 1] & $0.46^{+0.37}_{-0.36}$  &  $R_\mathrm{p}$  \\
$v_\mathrm{day-night}$   & [0, 20] & $4.43^{+1.80}_{-2.19}$  &  km\,s$^{-1}$  \\
$\epsilon$   & [0, 1] & $0.47^{+0.36}_{-0.33}$  &  -  \\
\hline
\end{tabular}
\end{table}

\twocolumn
\end{document}